# Role of Triplet-State Shelving in Organic Photovoltaics: Single-Chain Aggregates of Poly(3-hexylthiophene) versus Mesoscopic Multichain Aggregates

Florian Steiner, John M. Lupton, Jan Vogelsang*

Institut für Experimentelle und Angewandte Physik, Universität Regensburg, Universitätsstraße 31, 93053 Regensburg, Germany

**ABSTRACT:** Triplet excitons have been the focus of considerable attention with regards to the functioning of polymer solar cells, because these species are long-lived and quench subsequently generated singlet excitons in their vicinity. The role of triplets in poly(3-hexylthiophene) (P3HT) has been investigated extensively with contrary conclusions regarding their importance. We probe the various roles triplets can play in P3HT by analyzing the photoluminescence (PL) from isolated single-chain aggregates and multi-chain mesoscopic aggregates. Solvent vapor annealing allows deterministic growth of P3HT aggregates consisting of ~20 chains, which exhibit red-shifted and broadened PL compared to single-chain aggregates. The multi-chain aggregates exhibit a decrease of photon antibunching contrast compared to single-chain aggregates, implying rather weak interchain excitonic coupling and energy transfer. Nevertheless, the influence of triplet-quenching oxygen on PL and a photon correlation analysis of aggregate PL reveal that triplets are quenched by intermolecular interactions in the bulk state.

Triplet excitons play a major part in degradation processes and loss mechanisms in the bulk heterojunctions of organic photovoltaics.[1] For instance, triplet excitons can efficiently be quenched by ground-state triplet oxygen, which subsequently forms highly reactive singlet oxygen and irreversibly destroys the functionality of the active material.[2] In addition, singlet-triplet annihilation can limit the lifetime of singlet excitons in the organic material, which potentially also reduces the efficiency of organic solar cells.[3] For these reasons, aspects of the formation and lifetime of triplet excitons in organic semiconductor materials have received much attention, especially in poly(3-hexylthiophene) (P3HT), the structure of which is shown in Figure 1a.[2c,4] Initial measurements using photo-induced absorption showed that triplet states, T$_1$, are generated with a rate constant of $k_{isc}^{-1}$ = 1.2 ns and a lifetime of up to 77 µs.[4b] It was concluded that T$_1$ states play a significant role in singlet-triplet annihilation processes.[4b]

Subsequently, the nanoscale morphology was recognized to have a significant impact on $T_1$ states in P3HT.[4c] Although triplet excitons are rapidly generated in amorphous regions, no triplets are observed in highly-ordered (crystalline) regions.[4c] Recently, this observation was connected experimentally and theoretically to different electronic aggregation mechanisms, J- and H-type.[5] In contrast, single-molecule spectroscopy confirmed the presence of $T_1$ states, even in highly-ordered single-chain aggregates.[3a] The lifetime of $T_1$ was again found to be in the µs regime and, more importantly, these triplets were shown to be capable of quenching the complete photoluminescence (PL) of a ~100 kDa sized P3HT chain.[3a] These results imply a striking difference regarding the role of triplet excitons in single chains versus the bulk material.

Here, we resolve this divergence by pursuing the "molecular mesoscopic" approach, constructing bulk-like objects from single chains. We follow the formation of triplets by studying the evolution of fluorescence photon statistics from single chains to multi-chain aggregates. We start by preparing a sample with single isolated P3HT chains embedded in a poly(methyl-methacrylate) (PMMA) film. Under these conditions, single chains are known to form highly-ordered aggregates which display a high degree of polarization anisotropy in absorption and strong photon antibunching.[6] The positions of the single-chain aggregates are obtained by confocally exciting at 485 nm and ~0.5 kW/cm², detecting the PL with two avalanche photodiodes (APDs) after splitting the PL with a 50/50 beam splitter and scanning the sample (details of the setup and materials can be found in ref.[3a] and[7]). Diffraction-limited spots are observed, which are subsequently placed inside the excitation focus. Single-chain aggregate spectra are measured by rerouting the PL from the APDs to a spectrometer equipped with a CCD camera. The green curve in Figure 1d shows the summed up spectrum of 20 single P3HT chains and the green histogram displays the distribution in 0-0 peak positions of the spectra.[8] The measurement of the PL with two APDs in a Hanbury Brown-Twiss geometry allows one to calculate the intensity cross-correlation function, $g^{(2)}(\Delta\tau)$, which is obtained under pulsed excitation. The time difference $\Delta\tau$ between two consecutive photons is therefore measured in multiples of the difference in excitation laser pulse number $\Delta p$. Figure 1e displays such a $g^{(2)}(\Delta p)$ plot in which 50 single P3HT chains are summed up (green bars). Summation is carried out because one single P3HT chain does not emit sufficient photons to construct a telling $g^{(2)}(\Delta p)$ plot. The relative value of the center bar to the lateral bars relates to the probability of emitting a single photon at a time, denoted here as antibunching contrast, $A_c$.[9] A value of $A_c = 100$ % would correspond to a perfect single photon emitter and single P3HT chains embedded in PMMA come close with a value of $A_c = 76$ %. This value is a strong indication that efficient singlet-singlet annihilation and interchromophoric energy transfer occurs in a single-chain aggregate, similar as in multi-chromophoric model systems.[10]

In a next step, we employed solvent vapor annealing (SVA) on a sample, which contained ~20 times higher P3HT concentrations. A confocal fluorescence scan image of such a sample before SVA is shown in Figure 1b. SVA was performed for 30 minutes on the microscope setup using a gas-flow chamber with a mixture of 95 % acetone and 5 % chloroform vapor to induce aggregation.[11]

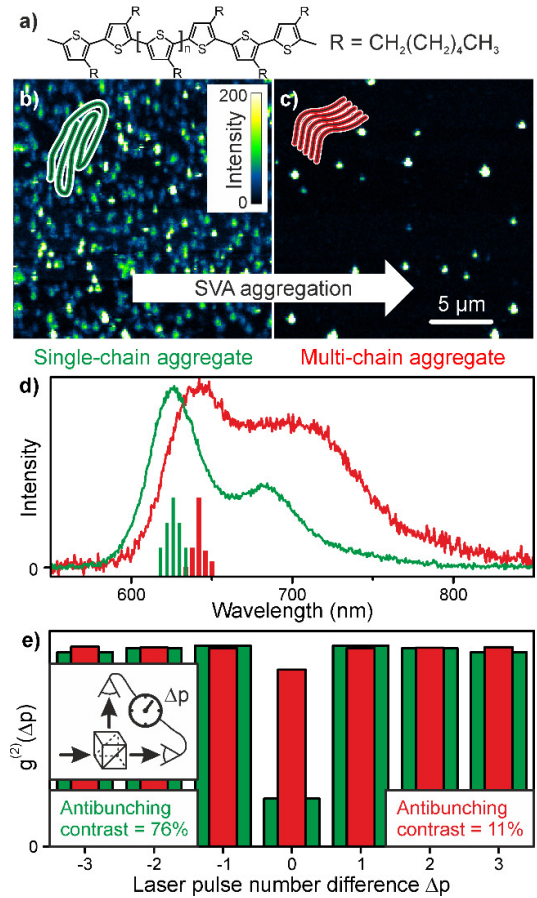

*Figure 1. (a) Chemical structure of P3HT. (b, c) Confocal scanning microscope images of P3HT embedded in PMMA before and after solvent vapor annealing (SVA). (d) Accumulated spectra of 20 single-chain (green line) and 50 multi-chain (red line) aggregates. Histograms of the 0-0 peak positions were constructed from the spectra of each species. (e) Accumulated intensity cross-correlation, $g^{(2)}(\Delta p)$, histograms for 50 single-chain (green) and 150 multi-chain aggregates (red). The single-chain and multi-chain aggregates were excited by laser pulses to control the difference in photon arrival time between the two detectors. The antibunching contrast is defined as the percentage by which the central bar at $\Delta p = 0$ is decreased compared to the lateral bars.*

Figure 1c displays the same sample after SVA with approximately one diffraction limited spot per 3×3 µm² area. By counting the number of spots per given area and comparing it with the sample before SVA, we estimate that one aggregate consists of ~20 single chains on average (see Figures S1 and S2 for a more detailed reasoning of this estimate[12]). The average spectrum of 50 aggregates and a histogram of the 0-0 peak positions of the individual spectra is given in Figure 1d (red). The aggregates exhibit significantly broadened and red-shifted PL spectrum with a decreased ratio between the 0-0 and 0-1 vibronic peaks compared to single-chain aggregates, i.e. similar to bulk-film spectra of P3HT.[13] With $A_c = 11\ \%$, the multi-chain aggregates do not show deterministic single-photon emission, implying that the aggregates are larger than the exciton diffusion length which determines whether two singlet excitons are present within a specific volume as required for singlet-singlet annihilation.

The impact of triplet excitons on the photophysics of conjugated polymers becomes directly apparent if the PL is measured in the presence of oxygen and in an oxygen-reduced environment.[14] Molecular oxygen effectively quenches the triplet, whereas in an oxygen-reduced environment, the triplet becomes long-lived with a lifetime ranging from µs up to seconds.[3a,14-15] Consequently, the PL intensity can be saturated under increased excitation densities due to shelving in the long-lived triplet.[14] Figure 2a-d displays confocal scan images of the same areas of single-chain aggregates (panels a and b) and multi-chain aggregates (panels c and d) under nitrogen and air. The sample was first measured under nitrogen and the same area was remeasured under ambient conditions by incorporating the sample in a home-built gas-flow chamber and purging this chamber with nitrogen or air. Selected diffraction-limited spots are marked with white circles for the different conditions. The PL intensity of single-chain aggregates increases by one order of magnitude by introducing air into the environment, as is apparent from the different intensity scales used for panels a and b. In contrast, by comparing panel c and d, it is concluded that the PL intensity of aggregates is independent of the atmosphere. This simple measurement establishes that triplets are indeed formed in P3HT and play a significant role in the photophysics of single-chain aggregates, but appear to be irrelevant in isolated multi-chain aggregates.

Two explanations of these observations are conceivable: (i) triplet excitons are also generated in multi-chain aggregates, but oxygen is not capable of quenching them as efficiently as in single-chain aggregates. For example, incorporation of dyes into supramolecular complexes can lead to reduced oxygen quenching and the same may apply to triplet excitons deeply embedded inside closely packed multi-chain aggregates.[16] (ii) Triplet excitons are not generated in the first place, or are removed very rapidly, for example by the triplet-polaron quenching mechanism.[17] For clarification, we resort to intensity cross-correlation analysis of the PL to reveal the presence of PL intermittency in the µs time regime, which is generally associated with the activation of triplets.[18]

Isolated diffraction-limited spots are placed inside the focus of excitation to record PL transients of single-chain and multi-chain aggregates. The insets of Figure 2e, f show examples of transients under nitrogen. The single-chain aggregate PL transient exhibits digital blinking on the time scale of seconds, and single-step bleaching after 34 s. This slow blinking behavior, which is also present in multi-chain aggregates, is primarily related to dynamic charging effects which have been previously examined extensively.[19] However, in this study only the triplets are of interest, with lifetimes in the µs regime, and for this reason we neglect the slow blinking kinetics by calculating $g^{(2)}(\Delta\tau)$ correlations for time delays, $\Delta\tau$, up to 1 ms. $g^{(2)}(\Delta\tau)$ correlations were calculated from 50 PL transients and the averaged correlation data (see Figure S3 for details on the averaging process) are shown in panel e. The correlation can be perfectly described by a single-exponential function (green line). The data reveal strong blinking kinetics in the µs time regime and the single-exponential fit suggests that only one dark state is involved, i.e. the

triplet.[3a] Analysis of the correlation function reveals a lifetime of ~18 µs for the dark state,[3a] consistent with literature.[4b]

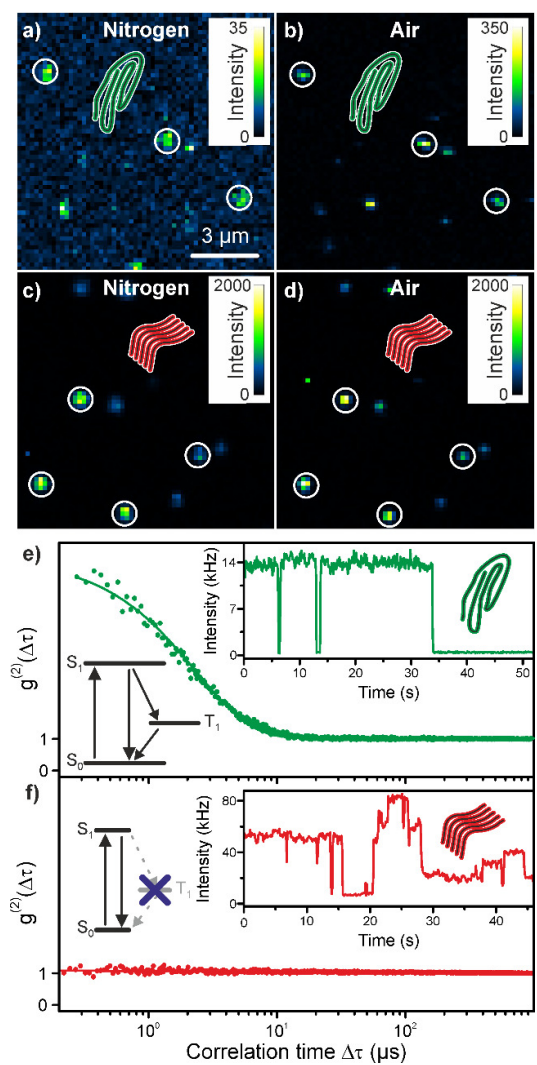

*Figure 2. Confocal scanning microscope images of single-chain (a, b) and multi-chain P3HT aggregates (c, d) measured under nitrogen (a, c) and in air (b, d). The same areas were measured for each sample. (e, f) Averaged intensity cross-correlation functions, $g^{(2)}(\Delta\tau)$, of 50 single-chain (e) and 150 multi-chain aggregates (f) under nitrogen atmosphere (see Figure S3 for averaging procedure). Energy level diagrams for single-chain and multi-chain aggregates are sketched. The insets display typical single-particle PL transients from which $g^{(2)}(\Delta\tau)$ was computed.*

The same dataset was obtained for 150 multi-chain aggregates (panel f) revealing a flat $g^{(2)}(\Delta\tau)$ curve at the Poisson limit of photon statistics (i.e., 1) under nitrogen atmosphere, indicating that no dark state is formed. However, there are two possible reasons for the absence of PL intermittency in the PL transients of aggregates: (i) Multiple independent emitters mask the impact of triplets. The average poor $A_c$ of 11 % suggests that at least 10 independent emitters are active in aggregates.[9] Only minor fluctuations in the overall PL intensity of ~11 % would then be anticipated under the assumption that a triplet impacts only the photophysics of the emitter on which it is formed. For this reason the $g^{(2)}(\Delta\tau)$ curve will stay close to the Poisson limit at all times. (ii) However, it is more interesting to hypothesize that triplets are either not formed or are quenched very rapidly in the multi-chain aggregates, so that there is not even any influence on the single emitter sites of the aggregate, the chromophores.

We must rely on the potential of sub-ensemble based techniques to test these hypotheses. Individual multi-chain aggregates exhibit sufficient photon counts to extract $A_c$. Figure 3 shows the distribution of $A_c$ values from the 150 measured multi-chain aggregates under nitrogen atmosphere. As expected from the mean $A_c$ of 11 %, most of the values range between 0 % and 10 %, but some outliers ($A_c > 40$ %) are measured with $A_c$ up to 66 %. These outliers are particularly interesting, because here we have a multi-chain aggregate with only 1-2 independent emitters, which can be analyzed regarding their triplet formation in the µs regime. It is important to note that these outlier aggregates are comprised of at least 4-15 single chains, based on the observed maximum PL intensity (see Figure S4). Three $g^{(2)}(\Delta\tau)$ curves ranging from $\Delta\tau = -10$ µs up to $\Delta\tau = +10$ µs are shown in the inset of Figure 3 for three different aggregates with $A_c = 6$ %, $A_c = 37$ % and $A_c = 66$ %. Crucially, no photon bunching due to singlet-triplet annihilation is observed on this time scale. This finding, in combination with high $A_c$ values of ~50 %, indeed suggests that no triplets are formed in multi-chain aggregates, or if they are formed, they are deactivated very quickly within the time resolution of the experiment (~25 ns, the repetition period of the laser). PL effectively arises from one single chromophore in this multi-chain aggregate, but this chromophore does not switch to a dark state or this dark state is rapidly quenched – it constitutes an effective three-level system with a non-detectable dark state (Figure 2e, f).

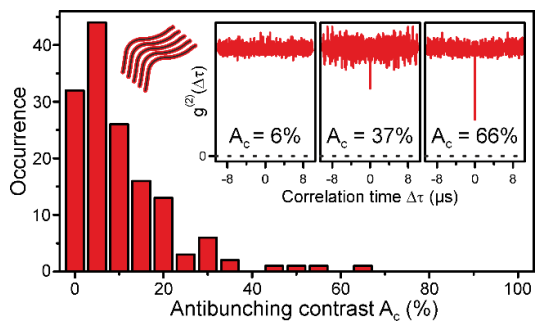

*Figure 3. Histogram of antibunching contrast, $A_c$, values extracted from intensity cross-correlation curves of multi-chain P3HT aggregates. Three examples of single-aggregate photon correlations are inset. Aggregates with a high $A_c$ value of ~66 % display no photon bunching in the µs time regime, indicating that triplets are not formed even though single-photon emission occurs.*

The type of excitonic coupling in P3HT is directly connected to the varying amount of intra- and interchain character and has been described within the weakly coupled aggregate model of Spano *et al.*[20] Recent work on different types of P3HT aggregates in the form of nanofibers brought to light that triplet formation is connected to the type of weak electronic aggregation, i.e. H-type or J-type.[5b] Nanofibers with H-type behavior appear to generate no triplets, in contrast to J-type nanofibers.[5b] Non-geminate recombination from long-lived polarons has been proposed to be responsible for the formation of triplet excitons in these J-type fibers.[5b,21] Comparison of the spectra of single-chain and multi-chain aggregates shown in Figure 1d with H- and J-type nanofibers of P3HT,[5b] suggests that the same mechanism is responsible here for the disappearance of triplets in multi-chain aggregates.

In conclusion, single-molecule and single-aggregate spectroscopy was employed to uncover distinct differences concerning triplet excitons in P3HT in the mesoscopic regime at the crucial transition from single chains to bulk materials. The work expresses both the strength and the inherent weakness of sub-ensemble-based techniques such as single-molecule spectroscopy. On the downside, important photophysical processes might be completely different in single chains and bulk films, leading to conflicting results reported for the same material. On the upside, close investigation and comparison of the photophysics from isolated single-chain and multi-chain aggregates reconciles these differences. The high yield of triplet formation in single P3HT chains along with the efficient quenching of singlet excitons by triplets suggests that triplet formation may limit the efficiency of organic photovoltaic devices.[3a] The fact that such quenching is manifestly absent in multi-chain aggregates implies that the problem is much less dramatic in bulk structures than previously anticipated.

## Supporting Information

Estimation of the multi-chain aggregate size from the PL intensity; averaging procedure of $g^{(2)}(\Delta\tau)$ in Figure 2e and f; estimation of multi-chain aggregate size displaying $A_c > 40\ \%$. This material is available free of charge via the Internet at http://pubs.acs.org.

## Corresponding Author

Jan.Vogelsang@physik.uni-regensburg.de

## Notes

The authors declare no competing financial interests.

## ACKNOWLEDGMENT

The authors are indebted to the European Research Council for funding through the Starting Grant MolMesON (305020).

# Supporting information for: "Role of Triplet-State Shelving in Organic Photovoltaics: Single-Chain Aggregates of Poly(3-hexylthiophene) versus Mesoscopic Multichain Aggregates"

Florian Steiner, John M. Lupton, Jan Vogelsang[*]

Institut für Experimentelle und Angewandte Physik, Universität Regensburg, Universitätsstrasse 31, 93053 Regensburg, Germany

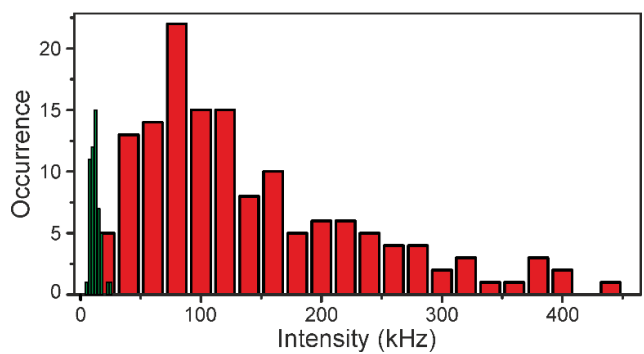

*Figure S1.* PL intensity histograms of the 150 multi-chain P3HT aggregates (red) shown in the main text and of 50 single-chain P3HT aggregates (green). The average PL intensity of the single-chain aggregates is 12 kHz, i.e. below the lowest intensity observed for a multi-chain aggregate. All samples were measured under similar experimental conditions. The multi-chain aggregates show increased PL intensity with an average value of 142 kHz and a maximum value of 445 kHz, i.e. on average the multi-chain aggregates are ~12× brighter than the single chains. We therefore conclude that the multi-chain aggregates consist of at least ~12 single chains, on average, and the largest aggregates may be comprised by as many as ~37 chains.

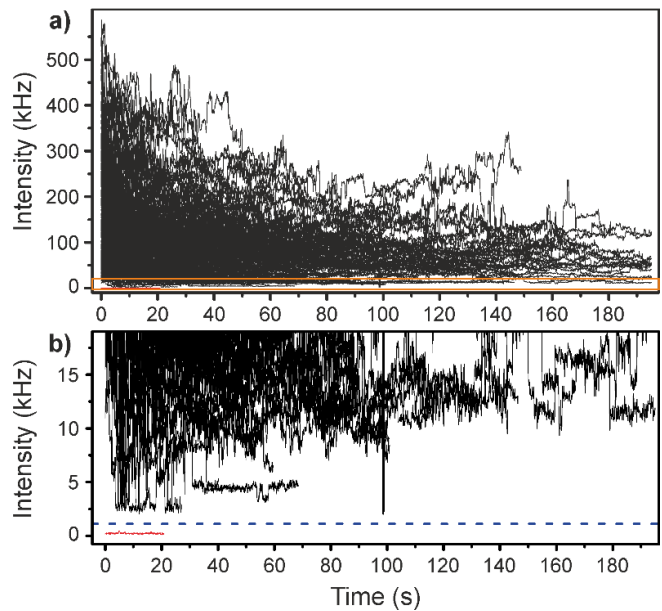

*Figure S2.* Summary plot of all PL intensity transients from the 150 multi-chain P3HT aggregates discussed in the main text. (a) The complete transients with a measurement time of 200 seconds are shown. (b) Close-up of panel (a) including a transient of the PL background (red) and the threshold limit applied (blue dashed line) to differentiate the presence of a multi-chain aggregate. All PL transients at all times during the measurement give count rates above the threshold limit, even if the PL is partially quenched by dynamic photoinduced charging events. We conclude that it is highly unlikely that a multi-chain aggregate would be missed in this analysis due to transient quenching and charging effects.[1] Therefore, the number of fluorescent spots before and after solvent vapor annealing is a good estimate for the average number of chains incorporated in a multi-chain aggregate, which yields ~20 chains per aggregate and is slightly higher compared to the minimum number extracted from the average PL intensity (see Figure S1).

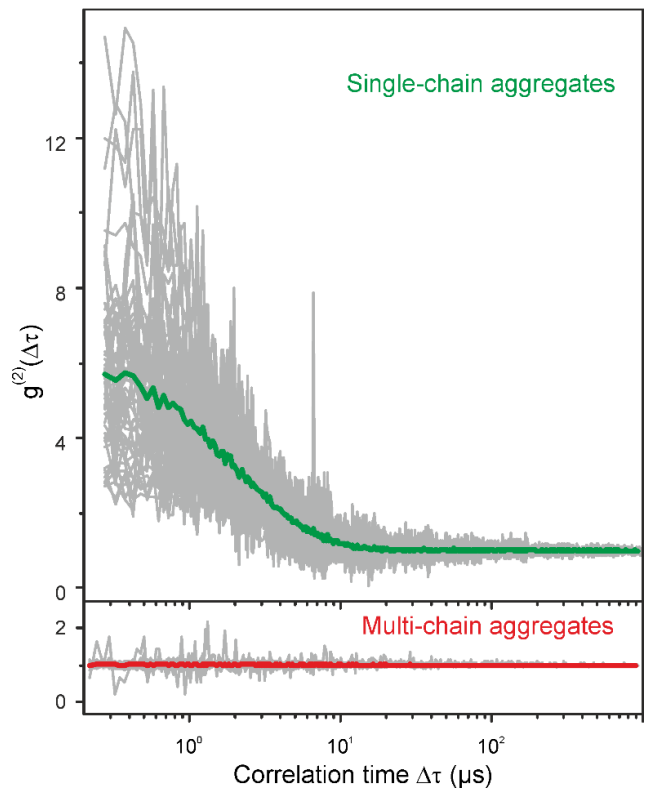

***Figure S3.*** *Averaged intensity cross-correlation functions, $g^{(2)}(\Delta\tau)$, of 50 single-chain (top, green curve) and 150 multi-chain aggregates (bottom, red curve) under nitrogen atmosphere. The $g^{(2)}(\Delta\tau)$ curves were first computed for all single PL transients (grey curves). However, these individual curves have an insufficient signal-to-noise level to extract lifetimes. In contrast, after measuring a sufficient number of single transients and averaging all $g^{(2)}(\Delta\tau)$ curves, the average can be used to extract a lifetime for the characteristic blinking kinetics, which is dominated by the triplet lifetime on these timescales. The analysis of the blinking kinetics was performed following the method outlined in ref. 3a of the main text.*

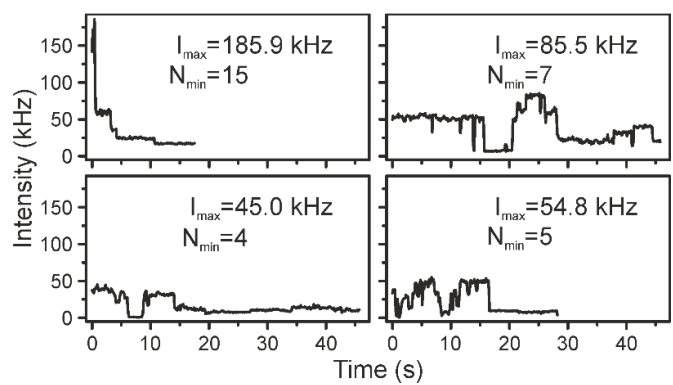

*Figure S4.* PL intensity transients of the four "outliers" shown in Figure 3, which exhibit an antibunching contrast, $A_c > 40\%$. The maximum PL intensity is stated in each panel and by comparing this value with the average PL intensity of a single-chain aggregate (12 kHz) the minimum number of single chains, $N_{min}$, incorporated in each aggregate can be extracted, as stated in the panels. We therefore conclude that these outlier PL traces cannot be due to single-chain aggregates.